\documentclass[times,twocolumn]{aastex62}

\usepackage[english]{babel}
\usepackage[utf8x]{inputenc}
\usepackage[T1]{fontenc}
\usepackage{amssymb}
\usepackage{multirow}
\usepackage{natbib}
\usepackage{afterpage}
\usepackage[caption=false]{subfig}
\usepackage[abs]{overpic}

\def\bi#1{\hbox{\boldmath{$#1$}}}

\newcommand{\mpc}{\ensuremath{\, h^{-1}\,\mathrm{Mpc}\, }}

\newcommand{\lya}{Ly$\alpha$}

\newcommand{\beq}{\begin{equation}}
\newcommand{\eeq}{\end{equation}}
\newcommand{\bc}{\begin{center}}
\newcommand{\ec}{\end{center}}
\newcommand{\bfig}{\begin{figure}}
\newcommand{\efig}{\end{figure}}

\usepackage{amsmath}
\usepackage{graphicx}
\usepackage[colorinlistoftodos]{todonotes}
\usepackage[colorlinks=true]{hyperref}
\usepackage{comment}

\begin{document}

\title{TARDIS Paper II: Synergistic Density Reconstruction from Lyman-$\alpha$ Forest and Spectroscopic Galaxy Surveys with \\ Applications to Protoclusters and the Cosmic Web}
\author{Benjamin Horowitz}
\affil{Princeton Department of Astrophysical Sciences, Princeton, NJ 08544, USA}
\affil{Lawrence Berkeley National Lab, 1 Cyclotron Road, Berkeley, CA 94720, USA}
\email{bhorowitz@berkeley.edu}

\author{Benjamin Zhang}
\affil{Department of Astronomy, University of California at Berkeley, Berkeley, CA 94720, USA}
\affil{Kavli IPMU (WPI), UTIAS, The University of Tokyo, Kashiwa, Chiba 277-8583, Japan}

\author{Khee-Gan Lee}
\affil{Kavli IPMU (WPI), UTIAS, The University of Tokyo, Kashiwa, Chiba 277-8583, Japan}

\author{Robin Kooistra}
\affil{Kavli IPMU (WPI), UTIAS, The University of Tokyo, Kashiwa, Chiba 277-8583, Japan}



\begin{abstract}

In this work we expand upon the Tomographic Absorption Reconstruction and Density Inference Scheme (TARDIS) in order to include multiple tracers while reconstructing matter density fields at Cosmic Noon ($z\sim 2-3$). In particular, we jointly reconstruct the underlying density field from simulated \lya\ forest observations at $z\sim 2.5$ and an overlapping galaxy survey. We find that these data are synergistic, with the Lyman Alpha forest providing reconstruction of low density regions and galaxy surveys tracing the density peaks. We find a more accurate power spectra reconstruction going to higher scales when fitting these two data-sets simultaneously than if using either one individually. When applied to cosmic web analysis, we find performing the joint analysis is equivalent to a Lyman Alpha survey with significantly increased sight-line spacing. Since we reconstruct the velocity field and matter field jointly, we demonstrate the ability to evolve the mock observed volume further to $z=0$, allowing us to create a rigorous definition of ``proto-cluster" as regions which will evolve into clusters. We apply our reconstructions to study protocluster structure and evolution, finding for realistic survey parameters we can provide accurate mass estimates of the $z \approx 2$ structures and their $z = 0$ fate.

\end{abstract}
\

\section{Introduction}

Starting with the discovery by \cite{gunn1965density} of the photo-ionized intergalactic medium (IGM), the absorption in the rest frame \lya\ by intervening neutral hydrogen in the spectra of luminous quasars, the so-called \lya\ forest, has given crucial insights to the high redshift universe. The power spectra of these fluctuations has been used to constrain exotic physics models \citep{2005PhRvD..71f3534V,2013PhRvD..88d3502V} as well as $\Lambda$CDM cosmology \citep{1996ApJ...457L..51H,2006JCAP...10..014S,2019A&A...629A..85D}, while individual features of the \lya\ flux\footnote{In this paper, the term `flux' implicitly refers to the \lya\ forest transmitted flux.} have been used to constrain reionization models \citep{2015MNRAS.447.3402B,2006AJ....132..117F} and find cosmic structures \citep{2015ApJ...814...40W,2017ApJ...837...71C}. 

In abstract, a sufficiently spatially dense collection of quasars at high redshift could be used to reconstruct the three dimensional map of the intergalactic medium, as first discussed in \citet{2001Pichon} and \citet{2008Caucci}. However, the relative rarity of luminous quasars have made making three dimensional maps of the flux field quite difficult (although see early attempts by \citealt{rollinde:2003} and \citealt{dodorico:2006}). Recently, there has been increased interest on tomographic reconstructions by also incorporating \lya\ forest absorption features observed in UV-emitting star-forming galaxies at z > 2, often referred to as ‘Lyman-Break Galaxies’
(LBGs). The observed forest flux features can be reconstructed in three dimensional space via a Wiener filtering algorithm \citep{2014LeeObserving} which can weight each sight-line based on its noise characteristics and impose a signal covariance based on the line of sight separations.


The ongoing COSMOS Lyman Alpha Mapping And Tomographic Observations (CLAMATO) survey \citep{Lee2017} has used 240 LBG sightlines covering a $\sim 600$ square arcmin footprint (a sightline density of 1455 $deg^{-2})$, reconstructing a 3D tomographic map of the $2.05<z<2.55$ \lya\ forest. This has allowed the probing of the three-dimensional (3D) structure of the optically thin IGM gas at $z>2$ on scales of several comoving Mpc. Another survey aimed at IGM tomography is the 
\lya\ Tomography IMACS Survey \citep[LATIS,][]{newman:2020}, which has observed a
wider area than CLAMATO, albeit at slightly lower spatial sampling.


Recently, \citet{Ravoux2020} carried out tomographic \lya\ forest reconstructions using 
data from the SDSS eBOSS survey. As opposed to CLAMATO, which used sightlines from both quasars and Lyman-break galaxies, the authors used only quasar sightlines from eBOSS. This resulted in a sightline density of 37 deg$^{-2}$, but on a much larger footprint of  220 deg$^2$. With a mean sightline separation of $13 \mpc$, the authors were unable to resolve individual features of the cosmic web, but were able to identify large protocluster candidates and voids.

The limited number of sight-lines within a given magnitude limit restricts the mean separation between sources and therefore the scales which can be reconstructed accurately. One possible approach to reconstruct smaller scales is to add additional constraints on the reconstruction beyond the assumed correlation function and 
spectroscopic noise properties used in direct Wiener Filtering. This was explored in \citet[][, hereafter TARDIS-I]{2019TARDIS}, where the authors created a dynamical forward model from the initial density field (at $z \sim \infty$) to the observed data. This was then cast as a Bayesian inference problem and optimized using standard solvers. Applying this technique to mock catalogs, TARDIS-I was found to outperform Wiener filtering both in terms of flux map reconstruction and overall cosmic web classification. However, as the \lya\ flux saturates in extremely dense environments, there was significant residual noise bias leading to an 
inability to accurately reconstruct protoclusters in the observed volume. \cite{2019porqueres} utilized a Hamiltonian Monte Carlo solver with a strong cosmological prior to reconstruct the density field with a similar foreword model.


One possibility to improve \lya\ forest reconstructions in overdense small-scale regions is to use additional information to jointly reconstruct the underlying density field. A natural choice for an additional probe is the observed 3D galaxy field overlapping the reconstruction volume, much of which will be collected incidental to any \lya\ tomographic spectroscopic survey. These galaxy populations are expected to be biased towards the more massive halos and provide a complementary probe of the cosmic structures.

When incorporating both galaxies and \lya\ flux within a dynamic forward modelling framework, one would expect synergy where both fields inform the reconstruction of each other. When viewed within the peak-background split framework \citep{1999MNRAS.308..119S}, we expect galaxies to fall in dark matter halos which will collapse preferentially in more dense environments (i.e. on the peak of long wavelength modes). This would mean that the positions of galaxies should inform the positions of more diffuse structures (as traced by the \lya\ forest) and visa versa. Alternatively, within a perturbation theory framework, one can view this as dynamical mode coupling \citep{1994ApJ...431..495J,2006PhRvD..73f3519C} which would allow one to infer the relative power of longer wavelength modes based on observations of short wavelength modes (and visa versa).

This joint reconstruction takes on additional significance due to upcoming surveys probing both the galaxy positions and \lya\ forest features over a wide sky area, including the Subaru Prime Focus Spectrograph Subaru-PFS \citep[PFS;][]{subaru} and Maunakea Spectroscopic Explorer
\citep[MSE;][]{2016MSE}. These telescopes are expected to have IGM tomography programs similar in noise properties and density to CLAMATO, but over a significantly wider area of sky ($\sim 50\times$ in the case of PFS, or 12-15 deg$^2$). The PFS program in particular is planned to specifically target galaxies in the coeval region with the tomography program in order to study the interplay between galaxy properties and cosmic structures (see the Appendix in \citealt{nagamine:2020}).

Farther into the future, thirty-meter class facilities such as the
Thirty Meter Telescope \citep[TMT;][]{2015TMT}, Giant Magellan Telescope \citep[GMT;][]{2012GMT}, and 
European Extremely Large Telescope \citep[EELT;][]{2014EELT}, have the potential to drastically improve sensitivity for faint background sources allowing for significant increases in sight line density and probe spatial scales of $\sim 1$ cMpc and below. The proposed 6.5m class MegaMapper facility \citep{2019BAAS...51g.229S}, on the other hand, would push into the full-sky regime to obtain spectra for $\sim 10^{8}$ galaxies between $z=2$ and $z=5$ over a hemisphere, which would immediately enable joint \lya\ tomography and galaxy analysis over $\sim$Gpc volumes. 

Performing this joint reconstruction within a forward modeling framework has a number of significant advantages for cosmological analysis. Assuming no primordial non-gaussianity (as suggested in the latest Planck cosmological analysis \citep{2018arXiv180706209P})  the power-spectrum of the initial density field should provide a lossless statistic, capturing all possible cosmological information. The entire family of higher order correlations and topological measures (such as three-point functions, density peak counts, voids, Minkowski functionals, etc.) arise due to gravitational evolution of a density field described by this primordial power spectrum. This encapsulation of a wide variety of cosmological probes has motivated a number of works in the field \citep[, e.g.,][]{seljak2017towards,2018Chirag}.

Beyond cosmological analysis, reconstructions from the \lya\ forest and galaxy field data can help inform the galaxy - cosmic structure relationship at high redshift. One of the original goals of the CLAMATO survey was to map out protoclusters and voids in their volume.  \citet{2015StarkProtocluster} pointed out, using N-body simulations and mock \lya\ forest
data, that the properties of high-redshift protoclusters observed with IGM tomography, such as their spatial extent,
can be linked to properties of their virialized $z=0$ cluster descendants (but see \citealt{cai:2016}, \citealt{cai:2017}, and \citealt{miller:2019}
for an alternative approach). This technique was applied
by \citet{2016LeeColossus} to study a $z\sim 2.4$ protoclusters detected in early CLAMATO
data. However, accurate initial density reconstructions from joint galaxy-\lya\ forest data
allow the possibility of directly modelling the gravitational evolution of \emph{individual}
observe high-redshift protoclusters all the way to $z=0$ and the fate of their constituent galaxies. Analogous techniques could also be used to study the evolution of individual void regions
detected in IGM tomographic maps \citep{Stark2015, 2018Krolewski}.

Direct comparisons between galaxy positions and their \lya\ forest environment have also 
been an active avenue for investigation in recent years \citep{font-ribera:2013, mukae:2017,bielby:2017,momose:2020a,momose:2020b}, in an effort to use the 
\lya\ forest flux as a proxy for the underlying density field. The assumption of
\lya\ flux as a density tracer is, however,
often complicated by hydrodynamical and radiative effects that can complicate the
relationship between matter density and \lya\ flux, as well as the fact that it is an 
exponential redshift-space tracer of the density field. Reconstruction methods like
TARDIS-I, on the other hand, offer a way to compare galaxy properties
with their surrounding large-scale ($>$1 Mpc) environments as defined directly by the matter
density field. At the same time they
also offer, at least in principle, a framework to simultaneously explore uncertainties in the flux-density mapping although we leave such investigations to future work.



In this work, we explore the possibilities of a joint reconstruction between an overlapping \lya\ forest and galaxy spectroscopic survey within the forward modelling framework.  In Section \ref{sec:method} we summarizes additions and changes to the technique described in \citep{2019TARDIS}. In Section \ref{sec:mocks}, we explain our mock generation procedure, in particular how we populate the density field with galaxies. We model our number densities and noise properties off of the upcoming Prime Focus Spectrograph. In  Section \ref{sec:results}, we show how the joint reconstructions improve the accuracy of the recovered density field and cosmic web classifications. In Appendix \ref{app:hydro}, we demonstrate that our reconstruction quality does not deteriorate appreciable when applied to mock catalogs including fully modeled hydrodynamical effects.

For our mock catalogs, we use the best fit cosmology from \cite{2016A&A...594A..13P}.

\section{Methodology}
\label{sec:method}

For a review of the optimization framework, see \citet[hereafter ``TARDIS-I'']{2019TARDIS}, as well as more general works \citep{seljak2017towards,2018Horowitz}. Here we specify how we expand our model to include galaxy fields and perform a joint optimization.

\subsection{Joint Likelihood}

The goal of our likelihood analysis is to maximize the combined likelihood of a prior term, a galaxy field term, and a \lya\ term. This can be expressed as likelihood on the initial density field, $\delta_i$ conditional on the observed flux, $\delta_{F,\textrm{obs}}$, and observed galaxy field, $\delta_{g,\textrm{obs}}$. This can be expressed as the sum of three individual log-likelihoods as 
\begin{eqnarray}
    \mathcal{L}_{comb}(\delta_i | \delta_{g,\textrm{obs}},\delta_{{\textrm \lya},\textrm{obs}}) =\mathcal{L}_{\textrm prior}(\delta_i) \nonumber \\
    + \mathcal{L}_{\textrm gal}(\delta_i | \delta_{g,\textrm{obs}})  + \mathcal{L}_{\textrm \lya}(\delta_i | \delta_{{\textrm \lya},\textrm{obs}}).
\end{eqnarray}
Here, we choose $\mathcal{L}_{\textrm prior}$ to enforce positivity of the powerspectra band-powers, but we do not impose any strong cosmological constraint on the fields. In full generality, we could also include a cross term between the galaxy field and the \lya\ forest field, for example if one expects strong feedback effects where the galaxies effect the observed flux field. At the scales of interest in this work ($\sim 1 \mpc$) there is no indication of a strong proximity effect nearby galaxies, but it could be included e.g.\ if known quasars
are in the volume (see \citealt{schmidt:2019}).

\subsubsection{\lya\ Likelihood}
\label{subsubsec:lya}
The \lya\ likelihood term is a comparison between the observed \lya\ forest flux and the reconstructed flux. This reconstructed flux is derived from the initial density field via a four step process, which is identical to that in TARDIS-I. The initial density field is evolved to the target redshift via flowPM, a GPU implementation of the fastPM framework \citep{fastPM} The evolved density is transformed to \lya\ optical depth, $\tau$, via the fluctuating Gunn Peterson approximation, $\tau = A \delta^\gamma$, with parameters $A=0.226$ and $\gamma = 1.5$. The optical depth is transformed to red-shift space using the evolved velocity field at the target redshift. The optical depth is transformed to flux, $F = \exp(-\tau)$, and the observed lines of sight are selected.

We can be expressed the log-likelihood in the Gaussian approximation as a chi-squared comparison of the observed flux with our forward modelled flux
\begin{equation}
    \mathcal{L}_{\textrm \lya}(\delta_i | \delta_{{\textrm \lya},\textrm{obs}}) = \sum_{n} \frac{(\delta_{{\textrm \lya},obs}(n) - \delta_{\delta_{{\textrm \lya},rec}}(n) )^2}{\sigma_{obs}(n)^2},
\end{equation}
where the sum over $n$ is over all individual pixel data and $\sigma_{obs}(n)$ is a noise level estimate from the spectra data reduction. We could also in full generality include off-diagonal elements in this likelihood to model additional effects such as continuum fitting errors. 

\subsubsection{Galaxy Likelihood}

For the reconstruction we want to roughly take into account the mapping from density to galaxies. To fully model this would require a differentiable (and fast) friends-of-friends halo finder and a framework to deal with the stochasticity of the galaxy population model. For simplicity, we will instead use a linear and quadratic bias term to model our real space galaxy field given the matter field at each step of the reconstruction. We will work in Lagrangian space , i.e.
\begin{equation}
    \delta_g(\bi{x}) = b_1 \delta_L(\bi{x}) + b_2^2 \delta_L^2(\bi{x}).
\end{equation}
The parameters $b_1$ and $b_2$ could be fit jointly, or calibrated via fitting the biased matter power spectra at some fiducial cosmology to the observed galaxy field. In addition, one could use a variant of Lagrangian halo bias estimators as explored in \citet{2017MNRAS.472.3959M}. In addition to this model, we explored a more standard Eulerian bias scheme, but found better agreement across a range of scales with a Lagrangian bias (see \citealt{2018galbias} for a thorough description of biasing schemes).

In analogy with the flux field described in Section \ref{subsubsec:lya}, we convolve the galaxies  with the peculiar velocity field to map to redshift space for comparison with the data. Here we only use objects with a spectroscopically confirmed redshift in our likelihood, and assume no error on that redshift. In practice, galaxy redshift errors ($\lesssim 100$ km s$^{-1}$ or $\lesssim 1\,\mpc$) from near-infrared spectroscopy are small compared to our forward model particle resolution thus we expect it to have minimal effect on our final reconstruction. 

We can compare our reconstructed galaxy field to the true galaxy field through a simple $\chi^2$ calculation with Poisson error-bars derived from the mock observed number counts. For a real survey, the noise covariance would also include a sizable contribution from the selection function, derivable from the spectrograph properties and survey strategy.





\subsection{Changes to Optimization and Forward Model}

We have also further optimized our reconstructions with the following additional changes and refinements in comparison with the method first presented in TARDIS-I.

\begin{enumerate}
    \item Our code has been completely rewritten in TensorFlow, allowing future easy integration with machine learning techniques and allowing optimization to utilize Graphical Processing Units (GPUs) for faster optimization. The underlying optimization is still performed with SciPy's LBFGS implementation.
    \item In order to implement our code on a GPU, we are using FlowPM for our dynamical forward model. This is a version of FastPM written in TensorFlow to allow automatic differentiation. 
    \item We have adjusted our optimization to perform a multiscale optimization where at each step in the optimization we smooth the resulting density field in steps, progressing from 5 Mpc/h smoothing to no smoothing. This is inspired by work done in \cite{2018Chirag}. The smoothing range we use was found through empirical testing to give the most accurate reconstruction.
    \item Changes in the reconstruction implementation have significantly reduced reconstruction time, allowing us to reconstruct larger volumes. In this work we analyze $L=128 \mpc$ cubes as opposed to $L=64 \mpc$ in \citet{2018Horowitz}, and each optimization takes approximately 10 minutes on a single NVIDIA Tesla V-100 GPU\footnote{Memory considerations limit larger volumes at the same particle resolution, but in practice large volumes can be reconstructed by concatenating smaller subvolumes with overlap to reduce edge effects.}.
\end{enumerate}

\begin{figure}
    \centering
    \includegraphics[width=0.45\textwidth]{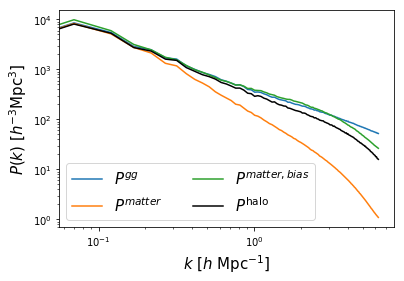}
    \caption{Calibration of galaxy bias by adjusting parameters $b_1$ and $b_2$ of the matter density field power spectra to match the observed galaxy field power spectra. While agreement breaks down at small scales, this is well below the $\sim 2.0 \mpc$ scale of interest in this work.}
    \label{fig:bias}
\end{figure}

\section{Mock Datasets}
\label{sec:mocks}



In this paper, we focus our reconstruction efforts on mock survey data that simulate the planned Galaxy Evolution Survey to be carried out as part of a planned Subaru Strategic Program (SSP) on the 8.2m Subaru Telescope's upcoming 2400-fiber Prime Focus Spectrograph \citep{sugai:2015}. The overall PFS Galaxy Evolution
Survey is planned\footnote{Note that while \citet{takada:2014} outlines an early version of
this survey, it is no longer up-to-date and pre-dates 
the PFS IGM tomography program.} to observe three separate extragalactic fields of $\sim 4-5\,\mathrm{deg}^2$ each, with 
multiple visits to build up $\sim 350,000$ medium-resolution spectra (covering 380nm-1.25$\mu$m) over multiple target classes ranging from $0.7\lesssim z \lesssim 6$. Of pertinence to us are the 
IGM tomography background and foreground components designed to cover 
the $2.2 < z < 2.7$ redshift range\footnote{See the Appendix in \citet{nagamine:2020} for more details on the PFS IGM tomography program.}, which we will aim to simulate in the subsequent mock reconstructions. 

\begin{figure*}[ht!]
    \centering
    \includegraphics[width=0.68\textwidth,clip=true, trim= 0 20 0 30]{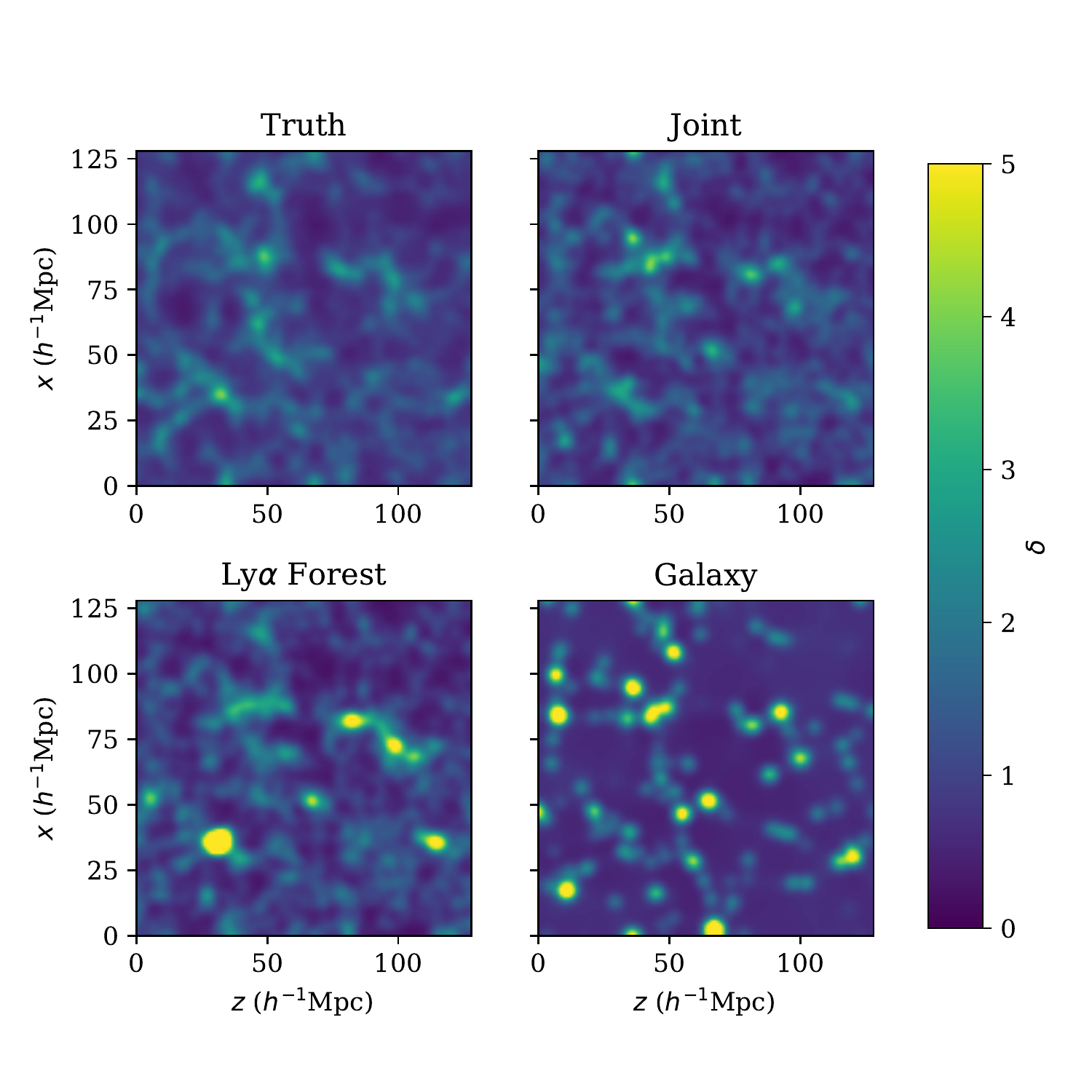}
    \caption{Average $z=2.5$ density field across a 10 \mpc slab projections through the $y$ axis for the three reconstructions, with the true density field for comparison. The joint reconstruction clearly yields a better match to the true density field than using the \lya\ forest and galaxy data individually. These density fields were smoothed at $2.0 \mpc$.}
    \label{fig:projected}
\end{figure*}

For the density field reconstructions, we run high resolution nbody simulations using a 10-step FastPM simulation with a particle resolution of $256^3$ over a (128 Mpc/h)$^3$ volume. FastPM has quickly become a standard tool for large mock catalog generation \citep[]{2019JCAP...09..024M} and has been demonstrated to reconstruct the statistical properties of other simulation methods with sufficient stage number \citep[]{fastPM}. As we will not be studying very small scale ($\lesssim$ 10 kpc) features within individual galaxy halos, fastPM provides a straightforward tool which can be easily interfaced to nbodykit to analyze the properties of group- and cluster-sized halos.

To populate our simulation with galaxies, we use the nbodykit friends-of-friends algorithm with linking length $0.2$ and minimum particle number of $5$ in order to find halos. We then use the \cite{2007Zheng} halo occupancy density model with $\alpha=0.5$ and $\sigma_{\log{M}} = 0.40$ to populate the halos with galaxies using halotools \citep[]{2017halotools}. From this catalog we can then model a realistic survey by uniformly down-sampling those galaxies, combining centrals and satellites with equal weight, to match the expected number density of our desired survey. We use the expected number density from the `IGM foreground' target class of the
Subaru PFS Galaxy Evolution Survey as our fiducial model, which corresponds to a projected density of 1000 galaxies per square degree coeval with the IGM tomography redshift of $2.1<z<2.6$. 
This is equivalent to a comoving number density of $n_\mathrm{gal} \approx 4.5\times 10^{-4}\;h^3\,\mathrm{Mpc}^{-3}$ within the redshift range. Within our $V=(128\,\mpc)^3$ simulation volume, we therefore assume a sample of 1000 foreground galaxies.


We then apply a redshift-space distortion along the line of sight based on the velocity of the underlying halos plus velocity dispersion. Note that at high redshift the relative impact of these distortions is small as most structure is still in the early stage of collapse and has relatively small line of sight velocity.

During our reconstruction we do not explicitly model individual galaxies but instead use them as a field. Computing the power spectra of the observed galaxy field, we approximately match the shape and amplitude of the matter power spectra with the choice of $b_1 = 0.301$ and $b_2 = 1.05$  (matching up to $k \approx 3.0$ ($h$ / Mpc)) for the matter field. We show this comparison in Figure \ref{fig:bias}.


We infer the hydrogen Ly$\alpha$ optical depth for the tomography using the Fluctuating Gunn Peterson Approximation (FGPA), with $T = T_0 (\rho/\bar{\rho})^{(\gamma - 1)}$ with slope $\gamma = 1.6$ \citep{2019TARDIS,lee:2015PDF}. This is then redshift space distorted and exponentiated to generate the underlying flux field. This approximation, as opposed to a fully hydrodynamical simulation, is justified in Appendix \ref{app:hydro} where the same reconstruction is performed on a hydrodynamical simulation. We do not use this field in this main work as we have found that the reconstruction quality does not appreciably suffer when using the fully hydrodynamical simulations, while the dark matter only n-body simulations allow direct comparison to previous works.

We then compute the flux field from the optical depth, $F = \exp(-\tau)$. The flux field varies from 0 to 1, with 0 being complete absorption and 1 being complete transmission of the background spectrum. We select random X-Y coordinate pairs, and take the flux along the Z-direction as one mock skewer. We select 2700 such skewers within our $L=128\mpc$ volume, in order to achieve the average sightline separation of the proposed PFS IGM tomography survey of $\sim 2.4$ $\mpc$. Note that it may seem counter-intuitive that the number of \lya\ forest skewers outnumbers the number of coeval galaxies (1000) within the same assumed survey, but this is due to the fact that each \lya\ forest sightline probes a redshift range of $\Delta z \sim 0.4-0.5$. Therefore, within our simulation box which corresponds to $\Delta z  = 0.14$ along the line-of-sight, \lya\ forest sampling comes from a relatively wide range of background source redshifts (see \citealt{Lee2014Theory}). 

We follow the procedure detailed in \cite{2019TARDIS} to add pixel noise to each mock skewer. In summary, each skewer is assigned a Gaussian noise level, which is constant across the skewer. To determine each skewer's noise level, we follow \citet{2015StarkProtocluster} and \citet{2018Krolewski}), by drawing an S/N ratio from a powerlaw distribution. Between a minimum and maximum S/N ($S/N_\mathrm{min}$ and $S/N_\mathrm{max}$), the distribution follows a power-law: $dn_\mathrm{skewer}/dS/N = S/N^\alpha$, with $\alpha = 2.7$. To replicate the noise properties of the Prime Focus Spectrograph, we then assign a minimum/maximum S/N of $S/N_\mathrm{min} = 2$, and $S/N_\mathrm{max} = 10$. Finally, we draw values for each pixel in the skewer from a Gaussian distribution with a standard deviation equaling the noise level, and add those values to the flux.

In order to model the error resulting from misfitting a background quasar or galaxy's spectral flux continuum, we apply the continuum error model of \cite{2018Krolewski} to the mock skewers. The flux values within each skewer is offset such that the final observed flux is
\begin{equation}
    F_{obs} = \frac{F}{1 + \delta_c},
\end{equation}
with $\delta_c$ being a value drawn from a Gaussian distribution with width
\begin{equation}
    \sigma_c = \frac{0.205}{S/N} + 0.015.
\end{equation}
We emphasize that this is a separate source of error from the pixel noise; $\sigma_c$ is added to the likelihood noise array $\sigma_{obs}$, but is not included in the pixel noise level used to draw noise for each pixel in a skewer.

\section{Results}
\label{sec:results}

We apply our updated TARDIS algorithm (hereafter ``TARDIS-II''), as described in Section \ref{sec:method} to the mock catalogs described in Section \ref{sec:mocks}. We explore a number of different properties of the reconstruction, both in terms of the mode reconstruction properties as well as in terms of $z = 2.4$ cosmic web classification. Since we are solving for the initial density fields that give rise to those structures, our reconstruction also will give us the velocity field at $z = 2.4$ and allow us to further evolve our particles to $z=0.0$.

\subsection{Density Reconstruction}
\label{subsec:dens_recon}

We qualitatively show the matter density fields of our different reconstructions in  Figure~\ref{fig:projected}. Visually, the \lya\ forest reconstructions does a good job of recovering the filamentary structures at close to mean density, but at overdensities the amplitude of the recovered structures is inaccurate. This is due to the known effect that the \lya\ forest saturates at relatively mild overdensities, therefore making overdensities ill-constrained. The galaxy-only reconstructions, in contrast, appear as peaks in the matter density field where galaxies trace strong overdensities, but does not recover more average densities and underdensities. The joint reconstructions using both \lya\ forest and galaxy information, on the other hand, yields both a good recovery of the overdensities as well the lower-density structures.

\begin{figure*}
    \centering
    \includegraphics[width=0.30\textwidth]{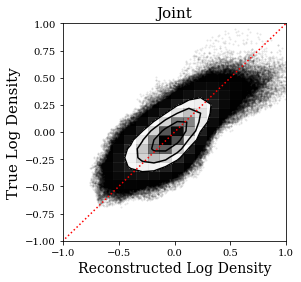}
    \includegraphics[width=0.30\textwidth]{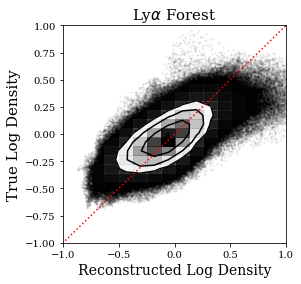}
    \includegraphics[width=0.30\textwidth]{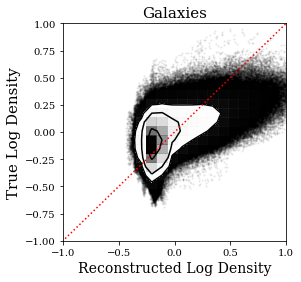}
    \caption{Scatter plot showing the reconstructed densities in the simulation volume in comparison with the the true density in each grid cell. The joint reconstruction clearly gives a more accurate and unbiased reconstruction of the density field compared with the \lya\ and galaxy reconstructions alone. All the densities have been smoothed on a 2 \mpc scale, which will be used for the later cosmic web analysis.}
    \label{fig:hist}
\end{figure*}

To assess the quality of our reconstructions, we first directly compare, on a cell-by-cell basis, the real space matter density field between the simulated truth and our reconstructions for each data combination. This is shown in Figure~\ref{fig:hist}. The joint reconstruction has significantly less bias and variance in the recovered matter field than either individual reconstructions. We can quantify the variance properties with the Pearson correlation coefficients, finding [0.72, 0.56, 0.54] for the joint, \lya\ forest, and galaxy reconstructions respectively ---  a coefficient of 1 would indicate a perfect
density reconstruction. The density fields in the joint reconstructions clearly have a more linear relationship with the true density field.

\begin{figure}
    \centering
    \includegraphics[width=0.4\textwidth,clip=true, trim=15 15 0 0]{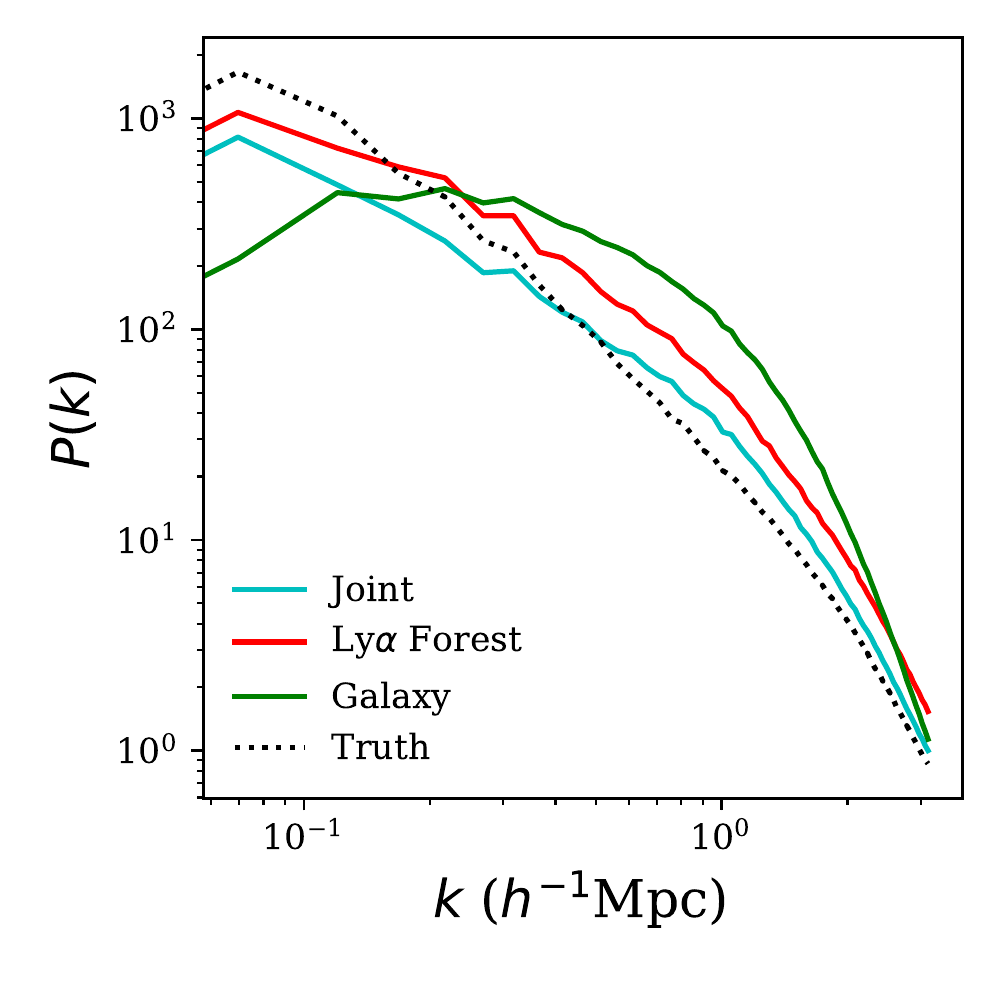}
    \caption{Power spectra of the reconstructed fields in comparison to the true power spectra at $z=2.5$. Note that while \lya\ forest and galaxies-only reconstructions have a significant excess of power at mid/small scales, the joint analysis has only a mild excess.}
    \label{fig:pk}
\end{figure}

\begin{figure}
    \centering
    \includegraphics[width=0.4\textwidth,clip=true, trim=15 15 0 0]{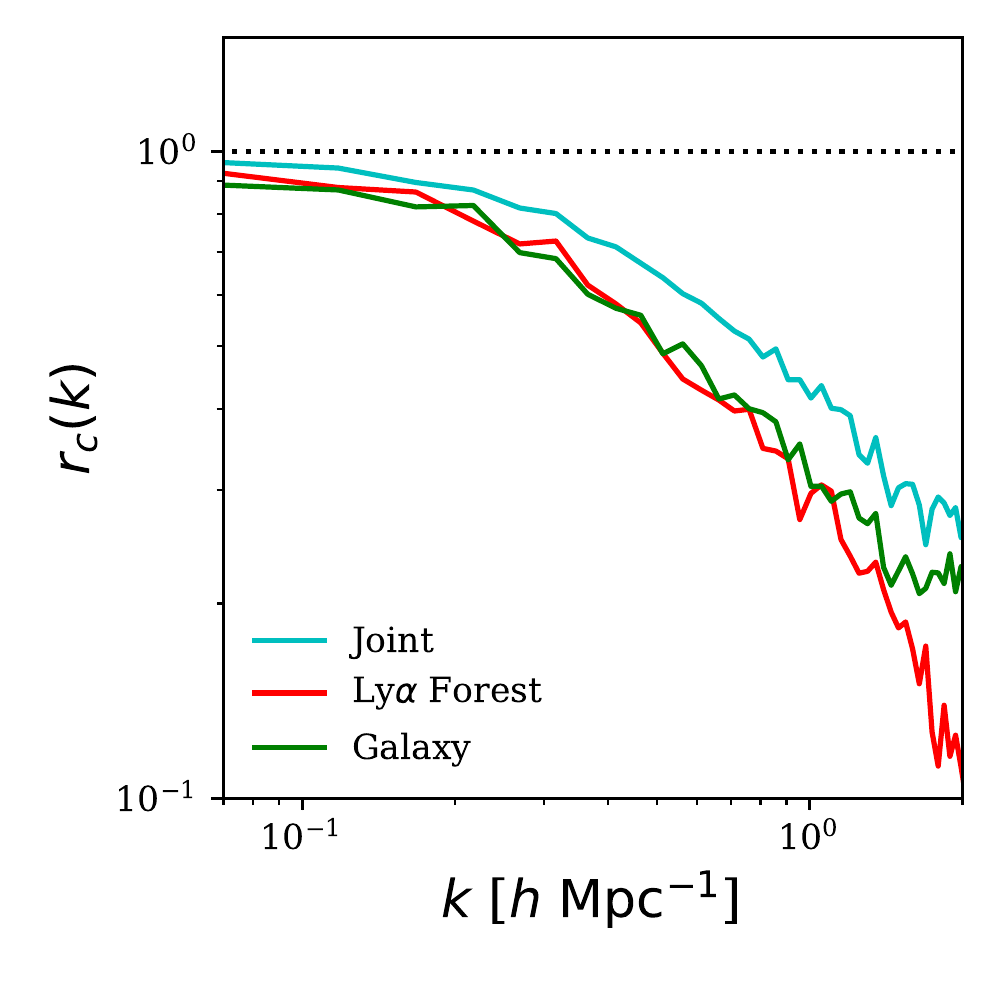}
    \caption{Correlation coefficient defined in Equation \ref{eq:crosscorr}, showing agreement between the two fields as a function of scale at $z=2.5$. Complete agreement would correspond to an $r_c = 1$. Each individual probe has a regime of scale with better reconstruction quality; galaxies appear to reconstruct small scale features while \lya\ forest is able to reconstruct the largest modes.}
    \label{fig:rc}
\end{figure}

To measure the quality of our reconstruction with the variety of datasets as a function of scale we use two-point statistics to quantify the fields. We use the reconstructed auto power spectra, $P_{RR}$, and 
cross-correlation coefficient defined in terms of the crosspower ($P_{RT}$) and auto-power ($P_{TT}$ for the truth) as;
\begin{equation}
    r_{c} = \frac{P_{RT}}{\sqrt{P_{RR}P_{TT}}}.\label{eq:crosscorr}
\end{equation}
We show the auto-power in Figure \ref{fig:pk} and for the cross-correlation coefficient in Figure \ref{fig:rc}. We find a comparable agreement across a range of scales, with the joint data-set having the best reconstruction across all scales, outperforming either probe individually. This suggests that there is a synergistic reconstruction, where our dynamic forward model is able to use the gravitational phase coupling information to inform the final optimized field.

Because TARDIS-II chooses a random Gaussian random field as its starting point, we expect to find some variance in a reconstruction for a particular dataset. In order to quantify the level of this variance, we run TARDIS-II 30 times on the mock catalogs with a random starting point, and calculate the cross-correlation coefficient for each run. We then consider the maximum difference in $r_c$ across runs for each scale, $\Delta r_c (k)$. For each individual probe, we find that $\Delta r_c$ remains at the $\sim 10^{-2}$ level across all scales. The joint dataset has a $\Delta r_c$ which is smaller than each individual probe, at the $\sim 5 \cdot 10^{-3}$ level. This suggests that the joint reconstruction has a synergistic effect for both reconstruction quality, and reconstruction variance.




\subsection{Cosmic Web Classification}

For a quantitative comparison of the cosmic web recovery in TARDIS-II, we use the eigenvalues and vectors of the pseudo-deformation tensor as described in \citet{2016LeeWhite,2017Krolewski} and inspired by work in \citet{Bond:1993,2007Hahn,forero-romero:2009}. This tensor has a strong physical interpretation within the Zel'dovich approximation \citep{1970Zeldovich} and, while there exist other cosmic web classification algorithms \citep[see summary in ][]{2014Cautun}, we choose to use the eigenvectors/values of the deformation tensor since it provides a continuous field which can be compared easily point-wise. Unlike past work, such as \citet{2016LeeWhite,2017Krolewski}, we will be working directly with the reconstructed density field rather than in flux. All the density fields used in our comparisons will be smoothed with a $R=2.0\,\mpc$ Gaussian kernel.

\begin{figure*}
    \centering
    \includegraphics[width=0.55\textwidth,clip=true, trim=20 0 0 0]{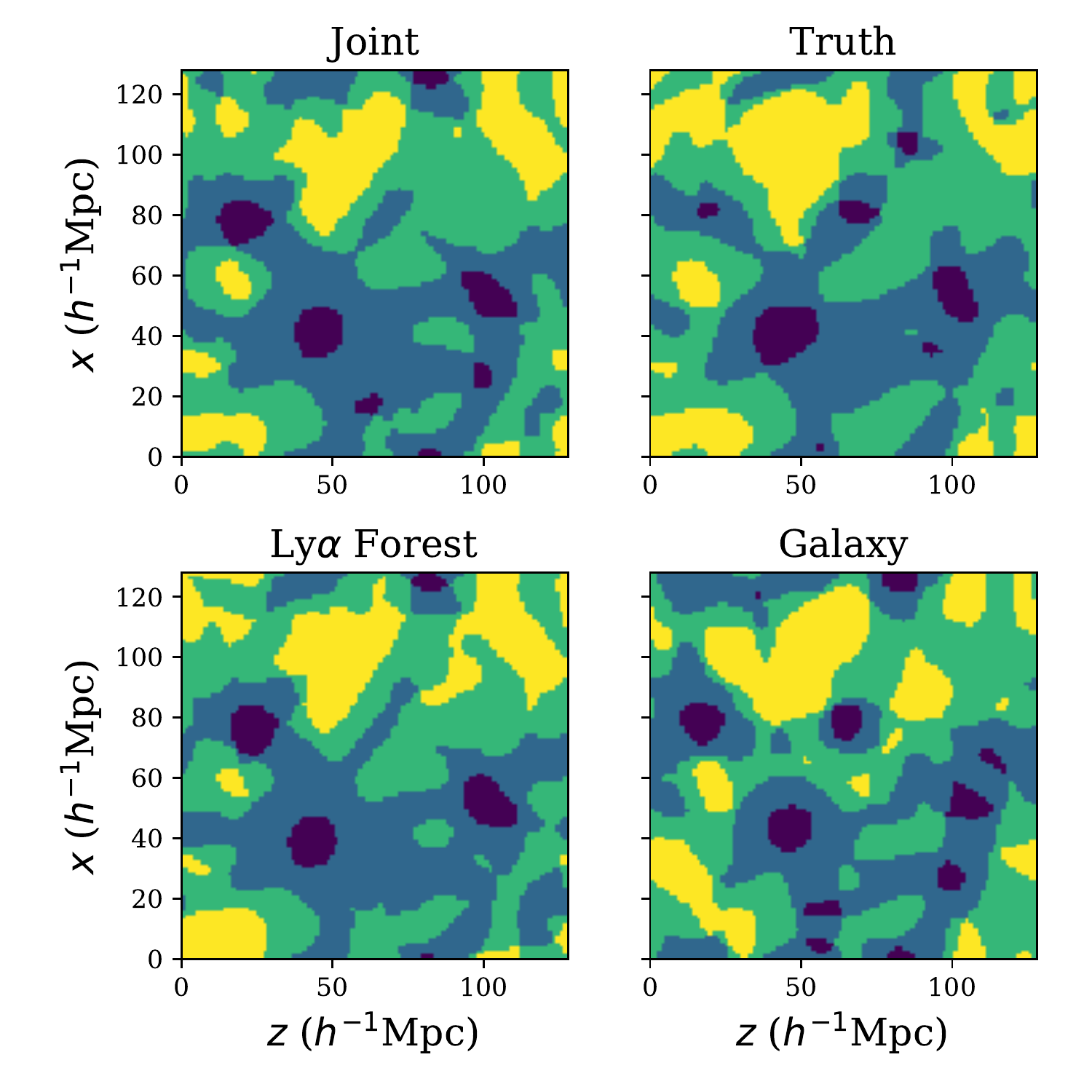}
    \caption{Cosmic Web Structures from a single 1 \mpc slice through our volume. 
    Dark blue indicates node, light blue indicates filament, green indicates sheet, and yellow indicates void regions. Qualitatively, the joint reconstruction better reconstructs the topology of the web.} 
    \label{fig:cosmic_web}
\end{figure*}

\begin{figure*}
    \centering
    \includegraphics[width=0.920\textwidth]{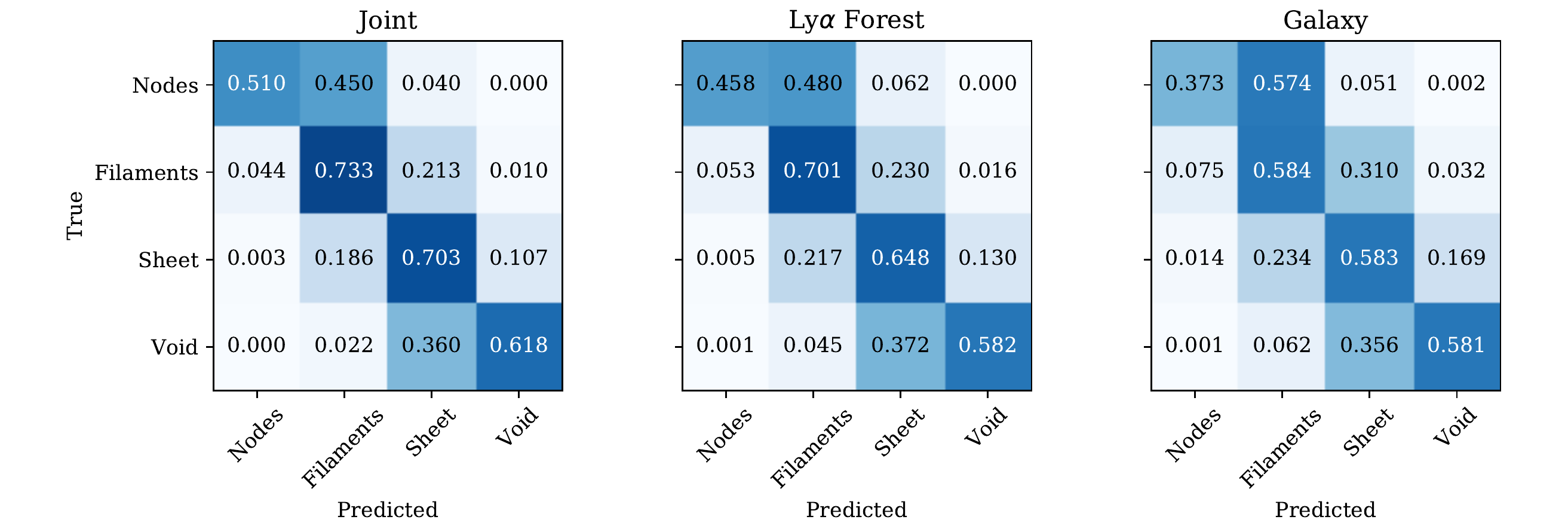}
    \caption{Confusion matrix for cosmic structures; numbers on the diagonal the fraction of each true cosmic structure correctly identified by the reconstruction, while those on the diagonal reflect the faction of each structure misidentified. The joint reconstruction appears to outperform either field individually across all cosmic structure types.}
    \label{fig:confusion}
\end{figure*}

\begin{deluxetable*}{@{\extracolsep{6pt}} l  c c c   c c c c @{}}
\tablecolumns{8}
\tablecaption{\label{tab:table1} Cosmic Web Recovery at $z=2.5$ (Eulerian Comparison) }
\tablehead{ \multirow{2}{*}{Mock Data}
   & \multicolumn{3}{c}{Pearson Coefficients} & \multicolumn{4}{c}{Volume Overlap (\%)} \\ \cline{2-4} \cline{5-8}
    \colhead{}  & \colhead{$\lambda_1$} &  \colhead{$\lambda_2$} & \colhead{$\lambda_3$} & 
      \colhead{Node} & \colhead{Filament} & \colhead{Sheet} & \colhead{Void} }
\startdata
        \texttt{Joint} & 0.853 & 0.796 & 0.708 & 51 & 73 & 70 & 62 \\
        \texttt{Ly$\alpha$}& 0.771 & 0.718 & 0.613 & 46 & 70 & 65 & 58 \\
        \texttt{Galaxy} & 0.746 & 0.635 & 0.518 & 37 & 58 & 58 & 58
    \enddata
  \label{table:1}
\end{deluxetable*}

The (pseudo-)deformation tensor is the Hessian of the underlying gravitational potential,

\begin{equation}
    D_{ij} = \frac{\partial^2 \Phi}{\partial x_{i} \partial x_{j}},
    \label{eq:diften}
\end{equation}
or equivalently in Fourier space in terms of the density field, $\delta_k$, as
\begin{equation}
    \tilde{D}_{ij} = \frac{k_i k_j}{k^2}\delta_k.
    \label{eq:diften_k}
\end{equation}

The eigenvectors of the deformation tensor relate to the principle curvature axes of the density field at each point, corresponding in the Zel'dovich approximation with the principle inflow/outflow directions. The corresponding eigenvalues determine if the net flow is inward or outward. Points with three eigenvalues above some nonzero threshold value $\lambda_{th}$ \citep[as in][]{forero-romero:2009} are classified as nodes (i.e. (proto)clusters), two values above $\lambda_{\rm th}$ are filaments, one value above $\lambda_{\rm th}$ are sheets, and zero values above $\lambda_{\rm th}$ are voids. The use of this thresh-hold value allows for the re-scaling of the relative number of each cosmic web type and allowing comparisons without an additional complication of overall normalization. Note that we could use the reconstructed velocity field to determine net inflow and outflow in a more physically accurate sense (i.e. beyond the Zel'dovich approximation), but keep the deformation tensor approach to maintain ease of comparison with other works (such as the Wiener filter). We follow past work \citep{2016LeeWhite,2017Krolewski,2019TARDIS} and define our threshhold value $\lambda_{th}$ for each reconstruction such that the voids occupy 21\% of the total volume at $z=2.5$ (see also \cite{2014Cautun}).

We calculate the deformation tensor on each reconstructed density field as well as the true density field, smoothed at 2.0 \mpc with a Gaussian kernel. We then choose a threshold $\lambda_{\rm th}$ for each reconstruction in order to keep the void fraction to 21\%. We show the qualitative results of this analysis in Figure \ref{fig:cosmic_web} for a given slice through the reconstruction, where one can see visually the improved reconstruction of cosmic structure with the joint analysis. A quantitative volume overlap comparison is shown in the confusion matrix in Figure \ref{fig:confusion}, as well as common misclassifications. When weighted by volume type, we find agreement of [68\%, 63\%, 58\%] for exact match in classification for Joint, Lyman-$\alpha$ forest, and Galaxy reconstructions, respectively. Significant misclassification, defined as classifications off by more than one eigenvalue sign (i.e. nodes mistaken as a sheet or void), are rare and comprise only [1.1\%, 2.0\%, 3.3\%] fraction of the volume for the joint, \lya\ forest and galaxy reconstructions respectively. 
This high-quality cosmic web reconstruction should allow PFS to confidently
study the variation of galaxy properties with respect to their cosmic web
environment at $z\sim 2.5$ (or lack thereof, e.g. \citealt{Martizzi2019}).

\begin{figure*}
    \centering
    \includegraphics[width=0.85\textwidth]{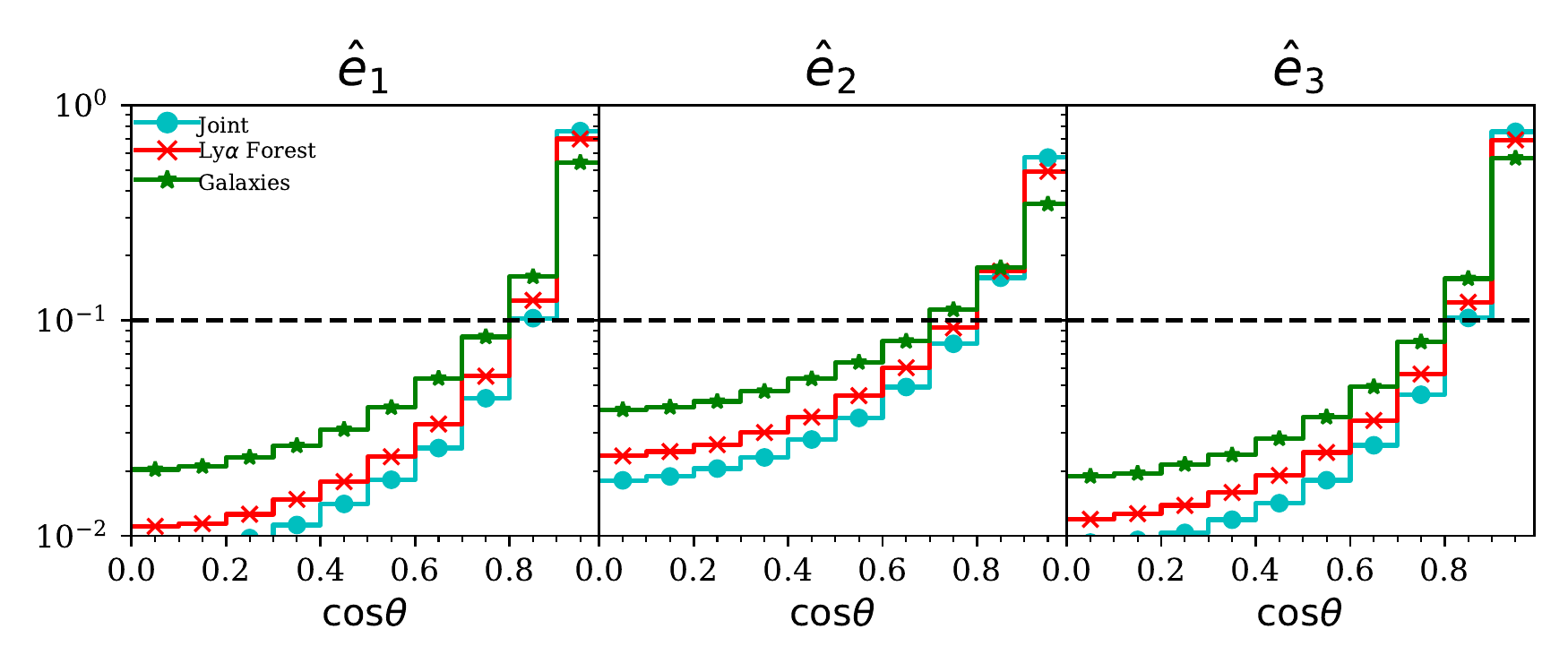}
    \caption{Cosine of the angle between the reconstructed eigenvectors and true eigenvectors of the density pseudo-deformation tensor. Random orientations would correspond to the black dotted line, while complete agreement would correspond to all density in the far right bin (i.e.\ $\cos\theta = 1$). The galaxy reconstructions perform considerably worse than the \lya\ forest since it doesn't properly recover filamentary structures, but the joint reconstruction significantly outperforms either probe individually.}
    \label{fig:eigenvectors}
\end{figure*}

In \citet{2019TARDIS}, we had studied the possibility of tracking the trajectories
of coeval $z\sim 2.5$ galaxies to their $z=0$ environments. While we do not
explicitly perform the same analysis here, on the basis of the superior cosmic web recovery 
of the joint reconstruction method we estimate that $\sim 50\%$ of
the $z\sim 2.5$ galaxies can be predicted to their correct $z=0$ environment.

We can also study the alignment of the reconstructed cosmic web with the true cosmic web via cell-by-cell comparison of the characteristic eigenvectors of the pseudo-deformation tensor. Matching the direction of the eigenvectors is particularly important for intrinsic alignment studies, where the direction of the filament is oriented along $\mathbf{\hat{e}}_3$ \citep[e.g.,][]{forero-romero:2014}.  Figure \ref{fig:eigenvectors} shows the dot product between the true and reconstructed eigenvectors, finding that the joint analysis provides substantial improvement in the recovery of the eigenvectors.
In $\sim 80\%$ of the grid cells, the recovered $\mathbf{\hat{e}}_1$ and $\mathbf{\hat{e}}_3$ are aligned to within $\cos\theta > 0.9$ of the true 
eigenvectors. This is comparable to the accuracy forecasted by \citet{2017Krolewski} for a TMT-like \lya\ forest-only survey using Wiener
filtering, although note that in this paper we actually adopt smaller smoothing scales ($2\,\mpc$ 
vs $4\,\mpc$). This therefore bodes well for the ability of Subaru PFS to
constrain intrinsic alignments between the galaxies and the
underlying cosmic web during the `Cosmic Noon' epoch of $z\sim 2-3$ \citep[e.g., ][]{codis:2018}, especially since the survey will deliver both the \lya\ forest sightlines
and foreground galaxy sample over a large volume.

\subsection{Cluster and Protocluster Reconstruction}

A common application of \lya\ tomography is the study of massive overdensities, such as galaxy (proto)clusters, at high redshift (as highlighted in works such as \citealt{Lee2017,2020ApJ...896...45M,2020ApJ...891..147N,Ravoux2020}). In this section we examine various statistics related to the reconstruction characteristics of the TARDIS-II technique to our mock catalogs.

\subsubsection{Cluster Reconstruction}

\begin{figure}
    \centering
    \includegraphics[width=0.45\textwidth]{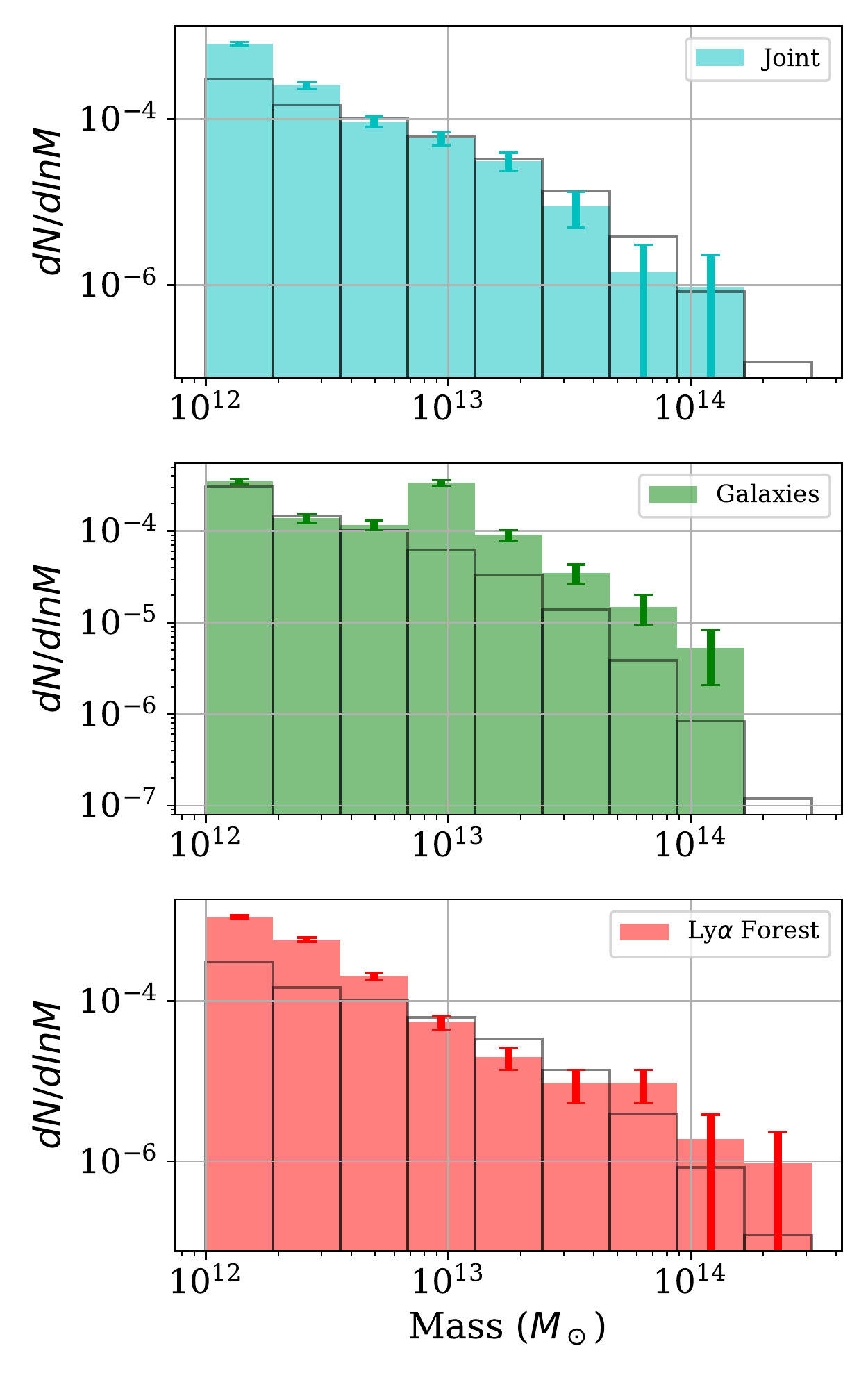}
    \caption{ Corresponding $z=2.5$ halo mass functions for the three reconstructions found via friends of friends halo finding. Black box represents the true halo mass function, while error bars are Poissonian. The joint reconstruction yields a halo mass function that is consistent with the true distribution at $M > 10^{12.5}\,M_\odot$.}
    \label{fig:hmf}
\end{figure}

\begin{figure}
    \centering
    \includegraphics[width=0.43\textwidth]{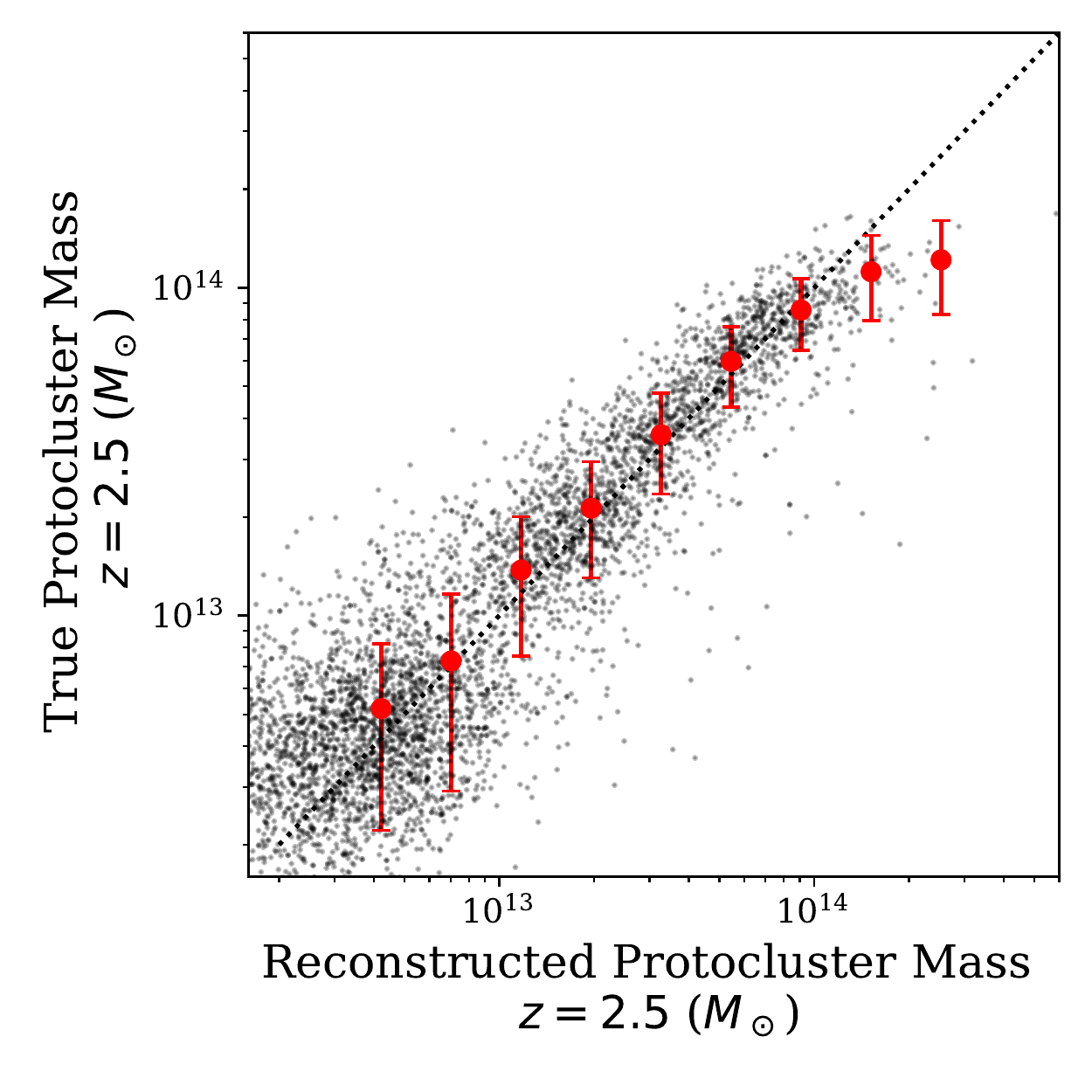}
    \caption{Comparison of the 5000 most massive identified $z=2.5$ protoclusters in the joint reconstructed volume, as identified by the predicted $z=0.0$ cluster positions, with the corresponding volume in the true field. Note that this mass is spherical overdensity mass, as opposed to FOF mass in Figure \ref{fig:hmf}.}
    \label{fig:individual_clusters}
\end{figure}
Since we have no prior on the underlying matter power spectrum, the lowest order statistic to check is the distributional qualities of the reconstructed massive halos in the form of a halo mass function. We use a friends-of-friends halo finder on our resulting particle positions at $z=0$ and $z=2.5$, with linking length of 0.2. In Figure \ref{fig:hmf}, we show the halo mass function for the three reconstructions in comparison to the truth. For the joint reconstruction we find the mass function is consistent with the truth at $M> 10^{12.5} M_\odot$, indicating very good agreement across a wide range of mass scales of astrophysical and cosmological interest, while for either galaxies or \lya\ alone there are significant residual biases. 

\begin{figure*}
    \centering\includegraphics[width=0.62\linewidth]{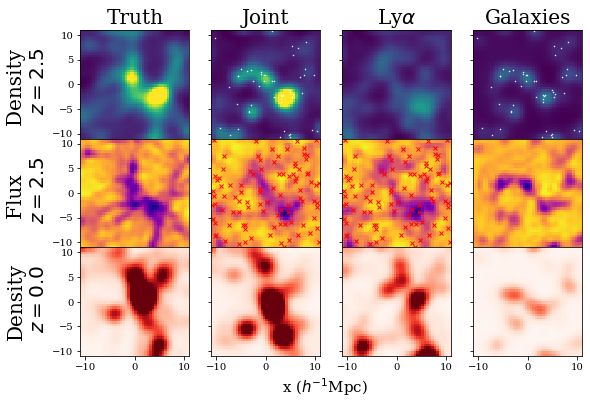}
      \newline

    \centering
    \textbf{(a)}  $M_{0} = 7.3 \times 10^{14} M_\odot$ cluster

    \centering\includegraphics[width=0.62\linewidth]{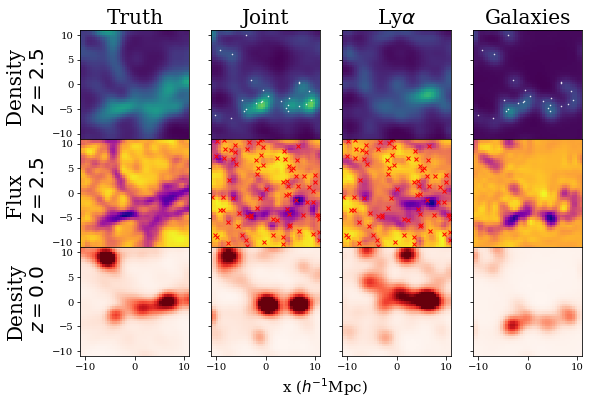}
      \newline

    \centering
    \textbf{(b)} $M_{0} = 6.2 \times 10^{13} M_\odot$ cluster


    \caption{Reconstructions of protocluster regions as identified by their predicted $z=0$ particle locations, projected over a 20 \mpc box in the line of sight direction.
    Clusters are categorized by their friends-of-friends halo mass at $z=0$, $M_0$, and the same Eulerian region of space is displayed at z=2.5 in flux and density. In the top row, the positions of observed galaxies are shown as white dots and \lya\ lines of sight are shown as red crosses. \textbf{(a)}; shows the most massive cluster in our mock volume, while \textbf{(b)} shows a more typical halo. }
    \label{fig:m1}
\end{figure*}

\subsubsection{Protocluster identification and reconstruction}

Although we are reconstructing the density and velocity field using $z = 2.5$ mock observables, we can further evolve the resulting fields to $z =0 $ to provide a rigorous definition of ``proto-cluster" as the progenitor
of modern-day galaxy clusters. 
In this case we will take any halo identified as a ``cluster" at $z=0.0$ by our friends-of-friends halo-finder and identify the corresponding particles at $z=2.5$ to define our proto-cluster region. We show projection plots of the most massive halos in our reconstructions in Figure \ref{fig:m1}.  These show the $z=2.5$ real-space dark matter density and redshift-space flux fields, as well as the evolved $z=0.0$ late time structure of these clusters. 
To compare masses of individual identified clusters at $z \sim 2.5$ we need to match analogous structures from our reconstructions and the simulated truth. While in abstract it is possible to perform this analysis with the friends of friends halo catalog used above, in practice there is some uncertainty due to the variation in the sub-structure within the halo causing slightly different halos to be identified (i.e. one large halo in the mock could be identified as two adjacent halos in the reconstruction, or vice-versa). To correct for this effect, we calculate protocluster mass by summing over the $z=2.5$ particle number within ten times the halo radius of each $z=0$ halo (i.e. an adjusted spherical over-density mass) at the $z=0$ Eulerian position. The choice of this radii was chosen to allow slight shifts in real-space position between the reconstructed and true halo positions. We show the protocluster mass recovery of the reconstruction in Figure \ref{fig:individual_clusters}. This indicates good agreement across a wide range of masses with little residual bias except at the very high mass end,
which suffers from small-number statistics. This plot could be roughly compared to Figure 9 in \cite{2016LeeColossus}, which included additional calibrations to determine a ``tomographic mass." They found roughly a factor 2 wider spread in calculated cluster mass compared to our
approach, although this did not include the galaxy field information.

\section{Conclusion}

\label{sec:conclusion}

In this work we have demonstrated the possibility of jointly reconstructing a density field from multiple disjoint tracers using a maximum likelihood framework. By combining a $z\sim 2.5$ \lya\ forest survey and an overlapping galaxy catalog that mock
up the upcoming Subaru PFS Galaxy Evolution Survey, we have shown that it is possible to reconstruct a more accurate 3D map than either one individually. We attribute this reconstruction quality to the multi-scale properties of our different unique tracers. Throughout all metrics we analyzed (power-spectra, cross-power, cosmic structure classification, and halo mass function) we found the joint optimization provides higher fidelity and less biased reconstructions. In light of upcoming surveys with overlapping \lya\ forest information and galaxy spectra, these joint analyses seem like a promising way to extract the optimal information.

In this work we use a simple biasing scheme to assign galaxies to the density field in our forward model. A more physical approach may be possible by using a more nuanced differential model to identify halos, such as a neural network as done in \cite{2019Modi}, and populate them according to a halo occupancy density model. This latter step is non trivial as it requires a differentiable stochastic model, but work in this direction is ongoing. However, it is useful to note that even with this extremely simple model we are able to achieve surprisingly accurate reconstructions of the cosmic web despite the sparse number densities used. In addition, more accurate object-by-object bias estimates could be extracted from spectra and photometry to better inform our reconstructions within the current framework.

Alternatively, instead of foreword modeling the HOD model, one could use a more nuanced biasing schemes such as that done in \cite{2019CosmicBirth,2020CosmicBirth}. In these works 
in these works a non-linear polynomial bias is used in Lagrangian space as well as a non-local tidal field bias encoded in the displacement fields to map the galaxy tracers from the Eulerian to the Lagrangian frame.
As long as the bias model is a differentiable function of the initial density field, it could be included in TARDIS. While not implemented in this work, it is also possible to jointly optimize for these bias parameters while solving for the initial density field.

Similarly, we assume the Fluctuating Gunn Peterson Approximation to model the \lya\ forest
in our simulations, which is known to break down in massive environments. \citep{sorini2018} While we explore the impact of these deviations briefly in Appendix \ref{app:hydro}, finding the effect on cosmic structure reconstruction small, for more futuristic potential surveys (i.e. TMT/GMT/EELT) it could be a significant source of error. In addition, a more detailed study of reconstructed cluster interiors would likely show significant systematic deviations when assuming FGPA. Depending on the application, a more nuanced model could be implemented to map from dark matter to hydrodynamical quantities; higher order analytical approximations, gradient based methods \citep{2018Dai}, or deep learning methods (i.e. \cite{2020arXiv200707267T}).


It is interesting to note that we are able to get high fidelity reconstruction quality through using an ``off the shelf" optimization scheme within SciPy's optimizer framework \citep[]{2020SciPy}. While in this work we did include an annealing scheme which improved reconstruction of large scale modes, this did not require any fine tuning to get satisfactory reconstruction quality, and wasn't used in TARDIS-I, the earlier iteration of our algorithm. Related techniques using Hamiltonian Monte Carlo for \lya\ flux reconstruction have also been developed \citep[]{2019porqueres,2020arXiv200512928P} but this approaches have significantly increased computational costs and it is currently unclear what benefits there are compared to an optimization approach for cosmic web recovery. 

These reconstructions also provide a possible powerful cosmological tool as we are reconstructing the initial density field corresponding the observed region which should contain all cosmological information. This aspect was a driving motivator for earlier interest in such reconstructions \citep[]{seljak2017towards}, and would be possible to do with our reconstructions as well. Residual biases in Figure \ref{fig:pk} could in theory be removed and associated covariance calculated within a response formalism (see \cite{seljak2017towards,2018Horowitz}). We leave this potentially promising avenue to future paper(s).

It is important to stress that the late time reconstructions are constrained to follow statistical properties of $\Lambda$CDM by nature of the forward model. For example, void density/velocity profiles would still be forced to follow the standard cosmological model. Deviations from this model (massive neutrinos, modified dark energy, etc.) would require alteration of the forward model. Either this deviation could be parametrized and fit jointly, or the reconstructions could be run separately with each model and the resulting likelihood values compared.

An additional natural extension to this work would be the inclusion of additional tracers beyond \lya\ forest and spectroscopic galaxy surveys, such as photometric redshift catalogs and cosmic microwave background lensing. The latter field could be of significant benefit, as the CMB lensing kernel peaks at $z \sim 2.0$ and it provides a direct probe of the matter density field on large scales. Not only could this provide a field complementary to both galaxies and \lya\, but it could be used to directly calibrate galaxy biases within the optimization. 

\section*{Acknowledgments}
We appreciate helpful discussions with Martin White, Zarija Lukic, and Chirg Modi. BH acknowledges support from NSF Graduate Research Fellowship (award number DGE 1106400) and JSPS (GROW program). This research was supported by the AI Accelerator program of the Schmidt Futures Foundation. Kavli IPMU was established by World Premier International Research Center Initiative (WPI), MEXT, Japan. KGL acknowledges support by JSPS KAKENHI Grant Numbers JP18H05868 and JP19K14755.

This research used resources of the National Energy Research Scientific Computing Center, a DOE Office of Science User Facility supported by the Office of Science of the U.S. Department of Energy under Contract No. DEC02-05CH11231.

\appendix

\section{Hydrodynamical Effects}
\label{app:hydro}

In the main body of the text we used the Flucuating Gunn Peterson Approximation for fast generation of mock catalogs and in our reconstruction. However, as the name suggests, this is only an approximation which is known to break down in various cosmic environments. In this section we briefly explore how well TARDIS performs on reconstructing density fields from Lyman Alpha forests including observational effects. We focus on the z=2.5 density field reconstructions on the map level. To perform this comparison, we will compare reconstruction quality using the full hydrodynamical flux field versus an FGPA transformed dark matter field for two sets of hydrodynamical simulations; NyX and Illustrius. How hydrodynamical effects propagate to our reconstruction quality can then be quantified by the degradation of each statistic. 

In order to isolate the hydrodynamical effects arising from feedback processes vs those arising from only baryon pressure, we use two different hydrodynamical simulations for the comparisons. The Nyx code \citep[]{2013nyx}, which relies on solving the evolution of a system of Lagrangian fluid elements coupled gravitationally to inviscid ideal fluid. The fluid evolution is modeled using a finite volume representation in an Eulerian framework using adaptive mesh refinement. While this technique allows accurate modeling of gas properties of diffuse structure (such as those which dominate the volume of \lya\ studies), it notably does not include feedback processes, such as AGN formation which could dominate at the centers of massive clusters. We use a 100 Mpc/h sidelength box with particle resolution $4096^3$ and choose our sightline and noise characteristics to match the PFS-like survey in the main text.  

\begin{deluxetable}{@{\extracolsep{6pt}} l  c c c   c c c c @{}}
\tablecolumns{8}
\tablecaption{\label{tab:nyx} Cosmic Web Recovery at $z=2.5$ (Variation of Forward Model)}
\tablehead{ \multirow{2}{*}{Mock Data}
   & \multicolumn{3}{c}{Pearson Coefficients} & \multicolumn{4}{c}{Volume Overlap (\%)} \\ \cline{2-4} \cline{5-8}
    \colhead{}  & \colhead{$\lambda_1$} &  \colhead{$\lambda_2$} & \colhead{$\lambda_3$} & 
      \colhead{Node} & \colhead{Filament} & \colhead{Sheet} & \colhead{Void} }
\startdata
        \texttt{Nyx} & 0.754 & 0.692 & 0.573 & 43.4 & 63.5 & 64.7 & 53.1 \\
        \texttt{FGPA}& 0.786 & 0.709 & 0.538 & 47.7 & 68.2 & 64.6 & 44.6 
    \enddata
\end{deluxetable}
\begin{deluxetable}{@{\extracolsep{6pt}} l  c c c   c c c c @{}}
\tablecolumns{8}
\centering
\tablecaption{\label{tab:ill} Cosmic Web Recovery at $z=2.5$ (Variation of Forward Model)}
\tablehead{ \multirow{2}{*}{Mock Data}
   & \multicolumn{3}{c}{Pearson Coefficients} & \multicolumn{4}{c}{Volume Overlap (\%)} \\ \cline{2-4} \cline{5-8}
    \colhead{}  & \colhead{$\lambda_1$} &  \colhead{$\lambda_2$} & \colhead{$\lambda_3$} & 
      \colhead{Node} & \colhead{Filament} & \colhead{Sheet} & \colhead{Void} }
\startdata
        \texttt{Illustris} & 0.739 & 0.689 & 0.595 & 42.4 & 62.3 & 63.9 & 47.7 \\
        \texttt{FGPA}& 0.794 & 0.744 & 0.629 & 47.5 & 66.7 & 67.2 & 50.6 
    \enddata
\end{deluxetable}
 
Unlike Nyx, the Illustris simulation does include feedback from AGN and star formation. For this purpose, we adopt the hydrodynamical Illustris simulation \citep{illustrisa,illustrisb,illustrisc,illustrisd}, which employs the adeptive mesh refinement code $AREPO$ \citep{arepo}. The AGN feedback prescription in Illustris produces too much heating, which is able to remove small-scale structure in the gas \citep[e.g.,][]{genel2014,park2018}. Therefore, it should provide a good test case here, since the difference from a non-feedback simulation such a Nyx would be relatively larger compared to other simulations with more realistic feedback prescriptions \citep[see e.g.,][for a detailed comparison between the two]{sorini2018}.

The dark matter density field from Illustris was deposited onto a grid using the same cloud-in-cell algorithm adopted in \citet{Martizzi2019} with 150 cells on a side and corresponding spatial resolution of 0.5 Mpc/h. 
We extracted skewers from the simulation along the z-direction with a resolution of 1 km/s on a regular grid separated 0.5 Mpc/h in the x- and y-directions. The skewers were generated using the $fake\_spectra$ package \citep{fakespectra}, which is designed to work on moving mesh simulations like Illustris to use the HI fraction directly from the simulation and takes into account the voronoi kernel adopted in $AREPO$. The Illustris-1 simulation volume has a sidelength of 75 Mpc/h and initially 1820$^3$ particles, resulting in a gas mass resolution of $1.3\times 10^6 M_\odot$ \citep{illustrispub}. 

To address how our reconstruction from quality degrades, we compare a reconstruction of using the hydrodynamical modelled optical depth for mock generation to a reconstruction using the Fluctuating Gunn-Peterson Approximation on the corresponding dark matter field for mock generation. This should allow us to compare in what sort of environments our approximate forward model fails.

\begin{figure}
\includegraphics[width=0.4\linewidth]{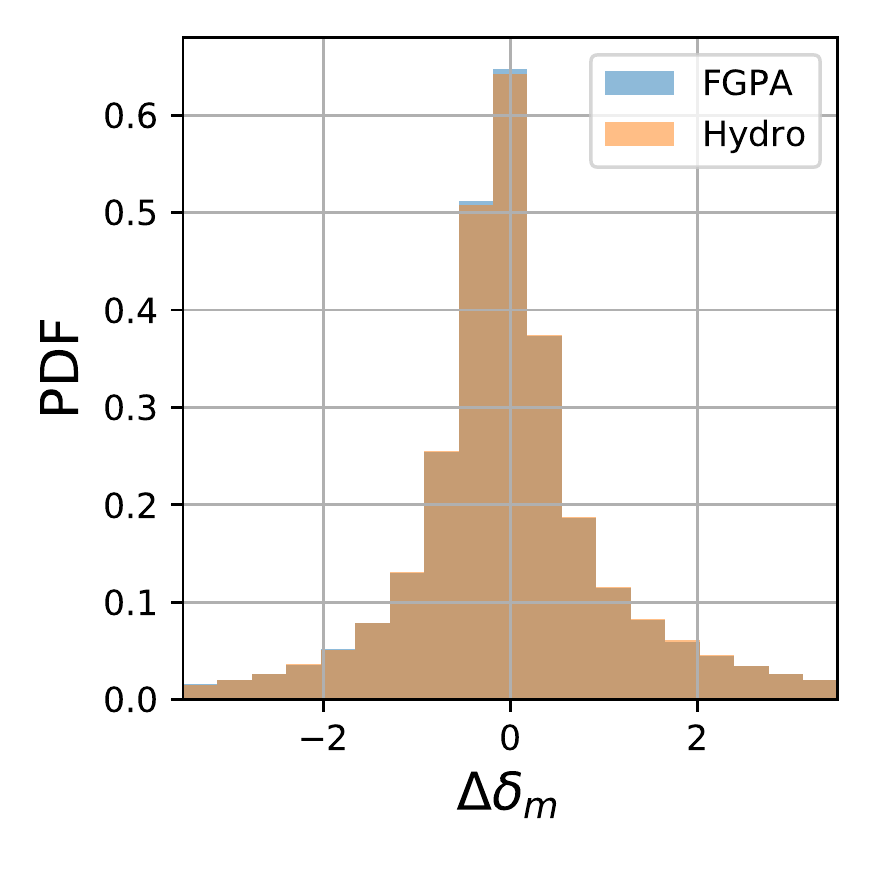}
\hspace{50pt}
\includegraphics[width=0.4\linewidth]{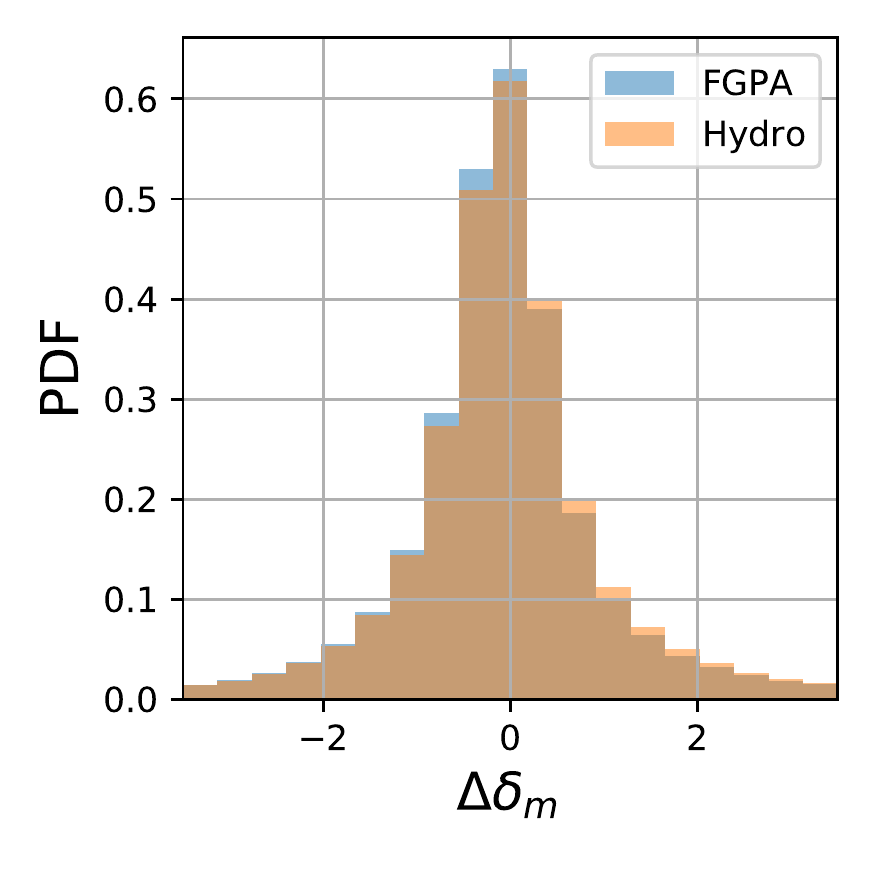}

\caption{Histogram comparing pixel error in reconstructed mass using mock catalogs generated off of Nyx (left) and Illustris (right) hydrodynamical models. The presence of feedback within Illustris slightly skews the reconstruction quality vs. those generated from dark matter alone.}
\label{app:flux}
\end{figure}

We show results for our reconstruction in terms of cosmic structure classifications for the $nyx$ simulation mock in Table \ref{tab:nyx} and for the Illustris simulation mock in Table \ref{tab:ill}, using the identical procedure as in the main text (i.e. comparable to Table \ref{tab:table1}). The performance of the two models in this context are quite comparable, with only a slight reduction in cosmic web characterization accuracy, justifying our use of the FGPA approximation in this context. In Figure \ref{app:flux} we show a comparison of the reconstructed density error for both simulation mock catalog reconstructions. In the absence of the feedback processes (i.e. for the $nyx$ simulation), the reconstructed density error is nearly identical whether or not hydrodynamical effects are included. With feedback, there is a slight skew in the reconstructed density, but still not sufficient to appreciably effect reconstructed cosmic structure.

\bibliographystyle{apj}

\bibliography{sample}

\end{document}